\documentclass[prb,twocolumn,letterpaper,citeautoscript,floatfix]{revtex4}
\usepackage{epsfig,subfigure,amssymb, amsmath, amsfonts,latexsym,array}

\newcommand{\pc}{$_\text{pc}$}

\graphicspath{{./figures/}}

\begin{document}

\title{Structural phases of strained LaAlO$_3$ driven by octahedral
tilt instabilities}
\author{Alison J. Hatt}
\author{Nicola A. Spaldin }
\affiliation{Materials Department, University of California, Santa Barbara}

\date{\today}

\begin{abstract}
We investigate the effect of epitaxial strain on [001]-oriented LaAlO$_3$ using
first-principles density functional calculations.
We find a series of structural phase transitions between states characterized by 
different patterns of tilting of the AlO$_6$ octahedra.
%
By tuning the biaxial strain from compressive to tensile, we induce an evolution in the
crystal structure in which the tilt axis changes from  
out-of-plane to in-plane, corresponding to space groups \em I4/mcm \rm and \em Imma\rm.
%
%
We also study the effect of uniaxial relaxation of the usual biaxial epitaxial
constraint and explore this as a mechanism for selectively stabilizing different
patterns of octahedral tilts.
\end{abstract}

\maketitle

\section{Introduction}

Oxides in the $AB$O$_3$ perovskite family present a multitude of functional
properties and are widely renowned for their potential in technological applications.  
Construction of heteroepitaxial thin films is being actively explored as a route to
further enhance and expand on the existing oxide functionalities. The presence of an interface
between an oxide and a substrate can dramatically affect material properties,
particularly if a film is grown coherently so that its in-plane lattice constant is
forced to match that of the substrate. The resulting heteroepitaxial
strain has been credited, for example, with dramatically enhancing the ferroelectric 
polarization and Curie temperature in thin film BaTiO$_3$ \cite{Choi_et_al:2004} and 
inducing ferroelectricity in usually non-polar SrTiO$_3$.\cite{Haeni_et_al:2004} 
In addition to ferroelectric distortions, we
increasingly find repercussions of strain in the patterns of rigid rotations of the
corner-sharing $B$O$_6$ octahedral units.   
These changes in rotation have been
invoked to account for such phenomena as the strain-dependence of magnetic properties,
as a result of the corresponding changes in the electronic bandwidth.\cite{Ishida/Liebsch:2008,Fuchs_et_al:2008}  
In this work, we use density functional theory to examine the effect of epitaxial strain
on the rotational instabilities in  LaAlO$_3$ (LAO).  
LAO is an ideal model system in which to isolate the
interplay between strain and rotations because 
it exhibits only rotational distortions in its ground state with no indications of
ferroelectric or Jahn-Teller instabilities.  

We are also motivated by a need to understand 
the influence of octahedral rotations in oxide heterostructures such as
LAO/SrTiO$_3$ (STO), which has been reported to form a
highly conductive electron gas at the interface in spite of the insulating nature of the
two parent compounds.\cite{Ohtomo/Hwang:2004} 
The propagation of octahedral rotations across the interface is likely to be relevant
to its electronic properties, but microscopic techniques have limited access to the
positions of oxygen ions.\cite{Jia_et_al:2009}
An improved understanding of the effect of strain on the parent compound LAO is
therefore highly desirable.

\begin{figure}
\centering
\epsfxsize=5.0cm \epsfbox{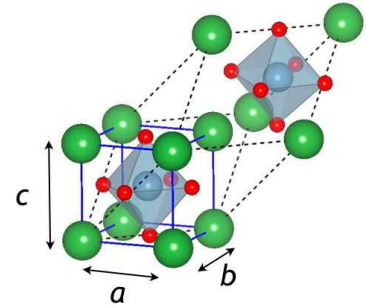}
\caption{Relationship between computational unit cell (dashed line) and pseudocubic
lattice parameters $a,b,c$ (solid blue line).  La atoms shown in green (dark grey); Al atoms
shown in blue (light grey) at center of oxygen octahedra.  (Color online.)
} 
\label{fig:uc}
\end{figure}

\begin{figure}
\centering
\epsfxsize=8.5cm \epsfbox{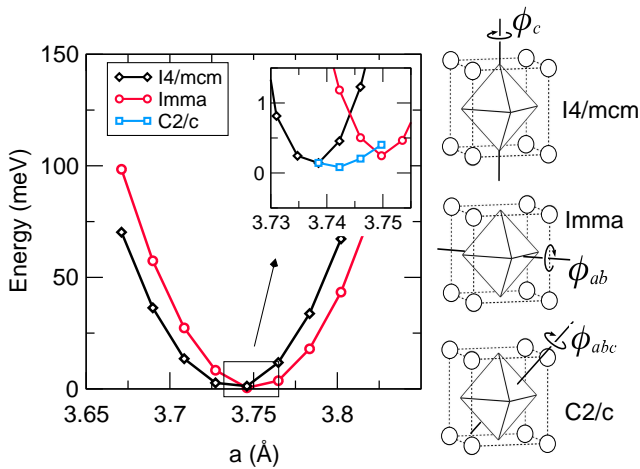}
\caption{
Energy per 5-atom formula unit of LAO as a function of in-plane
lattice constant $a$ (constant volume).
 Energy is given relative to unconstrained $R\bar{3}c$. 
{\sl Right:} Schematics of pseudocubic unit cells showing axes of octahedral
rotations in the three phases. (Color online.)
} 
\label{fig:strain-E}
\end{figure}

Here we investigate the effect of biaxial strain in LAO.  We find that the bulk
$R\bar{3}c$ structure is destabilized by the constraint of coherent epitaxy
and that compressive and tensile biaxial strains greater than $\pm$0.2\% stabilize
previously unidentified phases of LAO.
In addition, we find that uniaxial  relaxation of the biaxial epitaxial constraint
stabilizes a third phase. 
The primary structural differences between these strain-stabilized phases and the
parent phase are the patterns of octahedral rotations, and our analysis of the
transition between them provides insight into the coupling between strain and rotations
in LAO and similar $AB$O$_3$ perovskites.


\section{Calculation details and notation}

For all calculations we use the local density approximation (LDA) of density functional
theory as implemented in the Vienna \it ab-initio \rm simulation package,
{\sc vasp}.\cite{Kresse/Furthmuller:1996}  We use the projector augmented wave method
\cite{Blochl:1994,Kresse/Joubert:1999} with the default {\sc vasp} LDA potentials (La, Al,
O) and a plane wave energy cut-off of 800 eV. 
We use a 10-atom rhombohedral unit cell with a  5$\times$5$\times$5 {\it k}-point
sampling.
To determine ground state structures we relax ionic positions to a force tolerance of 
1 meV/{\AA}.
Space group determinations are performed with the {\sc findsym} symmetry analysis
software.\cite{FINDSYM}

Strain is calculated as $(a - a_0)/a_0$, where $a_0$ is the LDA equilibrium  lattice
parameter.  
We use the notation $a, b, c$ to denote the pseudocubic (pc) lattice constants (except
where noted), which correspond to the crystallographic directions [100]\pc, [010]\pc,
and [001]\pc.  
Our computational unit cell, shown in Fig.~\ref{fig:uc}, is defined by lattice vectors
($0,b,c$), ($a,0,c$), and ($a,b,0$). 
Lattice distortions comprised of rigid octahedral rotations are described by 
the angles of rotation around the pseudocubic axes, $\phi_a, \phi_b$, and $\phi_c$, as
shown in Fig.~\ref{fig:strain-E}. For completeness we also give the
Glazer notations for our structures, where, e.g.,  $\phi_a$ is written as
$a^+b^0c^0$ if consecutive octahedra along $a$ rotate in the same direction (in-phase)
or $a^-b^0c^0$ if consecutive octahedra rotate in the opposite directions
(out-of-phase).\cite{Glazer:1972}  
Multiple subscripts indicate a compound rotation, e.g. $\phi_{ab}$ denotes rotation 
around the $ab$ axis (crystallographic direction [110]\pc).
(Note that the $\phi_i$ are not strictly independent, however a rigorous decomposition
made using irreducible representations yields nearly identical results.)


\section{Biaxial  strain}

The bulk crystal structure of LAO deviates from the ideal $Pm\bar{3}m$ perovskite 
by out-of-phase rotations around the crystallographic [111] axis ($\phi_{abc}$, Glazer
system  $a^-a^-a^-$) that lower the symmetry to space group
$R\bar{3}c$, and an accompanying rhombohedral distortion. 
We reproduce this structure by relaxing lattice parameters and atom positions in our
10-atom unit cell, and
calculate a rhombohedral angle of 60.2$^\circ$ and an
equilibrium lattice constant of $5.298$ \AA\ (pseudocubic $a_0=3.746$ \AA), which underestimates the experimentally
determined value by $\sim$1\%, a common artifact of LDA calculations.   
We calculate $\phi_{abc}$ = 5.98$^\circ$, close to the experimentally determined
value.\cite{Muller/Berlinger/Waldner:1968} 

\begin{figure}
\centering
\epsfxsize=7.0cm \epsfbox{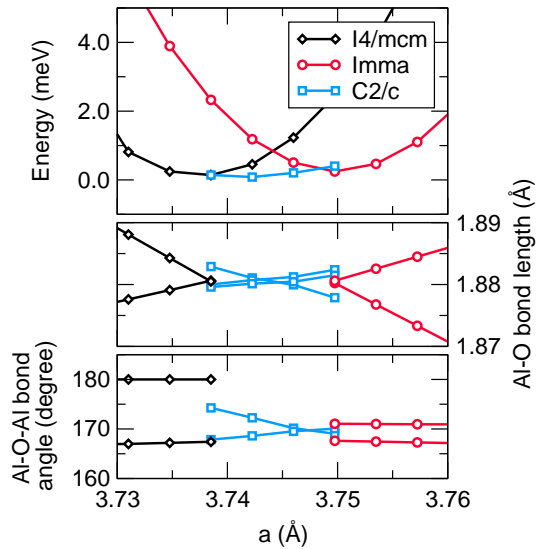}
\caption{Transition details  as a function of in-plane lattice constant $a$ (constant
volume).  
Top panel, energy per 5-atom formula unit; middle panel, Al-O bond distances; lower
panel, Al-O-Al bond angles.  (Color online.) 
}
\label{fig:order-params}
\end{figure}

\begin{table*}
\begin{tabular}{ m{2.2cm}  m{2.4cm}  m{2.2cm}  m{1.8cm}  m{2cm}  m{4.5cm}}
\hline\hline\noalign{\smallskip}
 & Rotation angles & & Rotation axis &Glazer notation& Conditions for stability\\
\noalign{\smallskip}
\hline\noalign{\smallskip}
$R\bar{3}c$ &\multicolumn{2}{m{4.4cm}}{$\phi_a=\phi_b=\phi_c\ne0$} &[111] &$a^-a^-a^-$ &
unconstrained\\
\noalign{\smallskip}
$C2/c$ &$\phi_a=\phi_b\ne0$ &$\phi_c\ne0$ &[111]\pc &$a^-a^-c^-$ &-0.2\% $< \eta
<$ 0.1\%, biaxial \\
\noalign{\smallskip}
$I4/mcm$ &$\phi_a=\phi_b=0$ &$\phi_c\ne0$ &[001]\pc &$a^0a^0c^-$ &$\eta <$
-0.2\%, biaxial  \\
\noalign{\smallskip}
$Imma$ &$\phi_a=\phi_b\ne0$ &$\phi_c=0$ &[110]\pc &$a^-a^-c^0$ &$\eta > $ 0.1\%, biaxial  \\
\noalign{\smallskip}
$Fmmm$	 &$\phi_a\ne0$ &$\phi_b=\phi_c=0$ &[100]\pc &$a^-b^0c^0$ & small uniaxial strain, $a > b$ \\
\noalign{\smallskip}
\hline\hline
\end{tabular}
\caption{Summary of the rotational modes and resulting space groups found in LAO under
various strain states. }
\label{table}
\end{table*}

To mimic clamping to a substrate, we enforce a square lattice in the plane of epitaxy
($a$ = $b$) and constrain the value of in-plane lattice parameter $a$ while relaxing the
out-of-plane parameter $c$ and atom positions.  We also allow a  [110]\pc\ shear of the
unit cell.  The
resulting structure, for $a=$ 3.746 \AA\, has space group $C2/c$; it   retains the
$\phi_{abc}$ rotations of the bulk and has a relaxed monoclinic angle
$\beta=$ 90.3$^\circ$. 
We then investigate the effect of biaxial strain by adjusting the value of $a$. 
For strains of -2\% to +2\% we identify two competing phases with space groups 
$I4/mcm$ and $Imma$.  
The resulting energy-versus-strain phase diagram is shown in Fig.~\ref{fig:strain-E}, and details
of the transition region are shown in Fig.~\ref{fig:order-params}.  
To facilitate fine-sampling near the transition we do not relax $c$ in these
calculations but instead maintain
the equilibrium volume.  Later we relax this constraint and find no qualitative changes
to the phase diagram.  We find that the shear distortion is only energy-lowering for the
$C2/c$ phase and all other phases retain $\beta=$ 90$^\circ$.

\begin{figure}
\centering
\epsfxsize=8.0cm \epsfbox{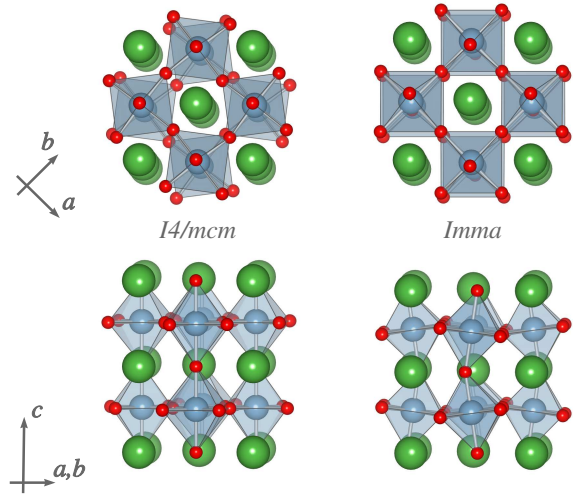}
\caption{Crystal structures of $I4/mcm$ (left) and $Imma$ (right) phases, the
ground-states for compressive- and tensile-strained LAO, respectively.  Upper figures
depict [001]\pc\ projection ($ab$ plane); lower figures depict [110]\pc\ projection.  La
atoms shown in green (dark grey); Al atoms shown in blue (light grey) at center of
oxygen octahedra. (Color online.)
}
\label{fig:structures}
\end{figure}

At very small strains of -0.2\% to 0.1\%, the ground state is $C2/c$.  The structure
exhibits the $\phi_{abc}$ rotations of the bulk and has a relaxed monoclinic angle
$\beta=$ 90.3$^\circ$.  As strain is increased, the $C2/c$ phase becomes rapidly unstable
and the system transitions to $Imma$ or $I4/mcm$, depending on the sign of the strain.
Compressive strain greater than -0.2\% stabilizes the $I4/mcm$ phase, in which the
lattice distortion consists of $\phi_c$ rotations (Glazer system $a^0a^0c^-$), whereas
tensile strain greater than 0.1\% stabilizes the $Imma$ phase, comprised of $\phi_{ab}$
rotations ($a^-a^-c^0$), as shown in Fig.~\ref{fig:structures}.  Thus, the bulk-like
$C2/c$ phase is found only in a very narrow range of strain near the equilibrium
lattice parameter.  These results are summarized in Table ~\ref{table} and structural
parameters for each phase at a representative strain are given in Table ~\ref{table2}.  (Note that the
precise range of stability for the three ground-state phases is difficult to identify
from total energies.  Thus we obtain our predicted range of the $C2/c$ phase by
calculating the zone-center phonons of each phase as a function of strain and comparing
the frequencies of the softest non-trivial modes.)

\begin{figure}
\centering
\epsfxsize=7.0cm \epsfbox{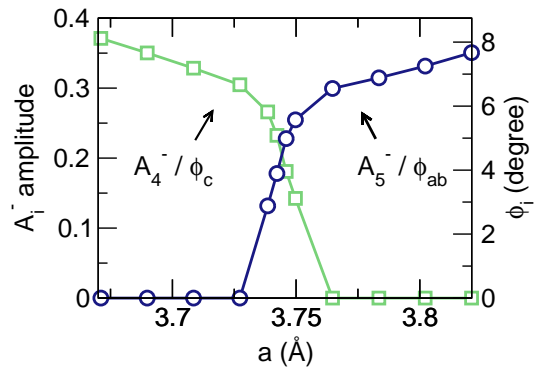} 
\caption{Rotation modes A$_i^-$ and equivalent tilt angles $\phi_j$ of LAO as a function of in-plane lattice parameter $a$ (constant volume).
(Color online.)
 }
\label{fig:rot_angle}
\end{figure}

To examine the coupling between strain and rotations, we decompose the distortion from a
high-symmetry parent
phase with space group $P4/mmm$ in terms of irreducible representations (or {\em
irreps}) using the software {\sc isodisplace}.\cite{Campbell_et_al:2006}
We find that the dominant irreps are $A_4^-$ and $A_5^-$, which correspond to $\phi_c$
and $\phi_{ab}$ rotations, respectively.  The evolution of the irrep amplitudes in the
lowest-energy structures is shown in Fig.~\ref{fig:rot_angle}.  
We also calculate  $\phi_c$ and $\phi_{ab}$ from the angles
between neighboring octahedra (in the manner of Ref.~\onlinecite{Zayak_et_al:2006}) to
provide an approximate correspondence between irrep amplitude and rotation angle
(Fig.~\ref{fig:rot_angle}).  Within $I4/mcm$ and $Imma$ we
find a roughly linear dependence of rotation angles on lattice parameter. 



Our results suggest a simple model to describe the response of LAO to biaxial strain.
In bulk LAO the $\phi_{abc}$ rotations reduce the bonding distances around the
small La cation and minimize the sum of Coulombic and repulsive
energies, while the dimensions of the octahedra are determined by the strongly covalent
Al-O bonds.\cite{Woodward:1997}
Within this lattice, strain-related changes in lattice spacing are accommodated
by changes in either the rotations or the Al-O bond lengths.  Under biaxial strain, we
find two distinct
responses within the range of strains investigated, as seen in the evolution of Al-O
distances and octahedral rotations, shown in Fig.~\ref{fig:order-params}.
At very small values of strain (-0.2\% to 0.1\%) the changes in lattice dimensions are
accommodated primarily by rotations while Al-O distances remain nearly constant, and the
system remains in the $C2/c$ phase.
Under compressive (tensile) strain, transformation to $I4/mcm$ ($Imma$) occurs when
$\phi_{ab}$ ($\phi_c$) goes to zero.
Increased strain is then accommodated by changes to Al-O distances, while the rotations remain nearly constant.  
We note that the precise values of strain at which the phase transformations are
predicted to occur are dependent on very small energy differences and are therefore
somewhat sensitive to the computational parameters used.  For example, test calculations
performed using the generalized gradient approximation for the exchange-correlation
energy indicated small variations in the quantitative details of the phase diagram but
the qualitative model does not change significantly. 


Similar strain-induced variations in tilt patterns have recently been
reported in a number of other perovskite oxides, including 
SrRuO$_3$,\cite{Zayak_et_al:2006} BiFeO$_3$,\cite{Hatt/Spaldin/Ederer:2010} and
LaNiO$_3$.\cite{May_et_al:2010}
The behavior of LAO is distinct from these systems, however, in the extremely narrow
window in which the parent phase is stable and the dramatically 
different structures stabilized by compressive and tensile strain.


\section{Uniaxial  relaxation of strain}

We next examine the effect of a uniaxial relaxation of tensile strain in LAO. 
We compare the energetics of $\phi_{ab}$ rotations ($Imma$ phase) to a phase with only
$\phi_a$ rotations (space group $Fmmm$).  
Under biaxial strain, the $Fmmm$ phase is $\sim$0.5 meV higher in energy than $Imma$ in
the range of strains investigated,
but a uniaxial relaxation of the strain, such that $a \ne b$, alters the balance of
energy between the two.
Starting with atom configurations from the relaxed structures for $a=b$, we vary 
the ratio $a/b$ and calculate the total energy for the two phases. 
We maintain a constant in-plane area $ab$ and do not relax $c$. 
We first explore the system with $ab=$ (3.746 \AA)$^2$ and find that   
a 0.5\% distortion of $a/b$ stabilizes $\phi_a$ rotations relative to $\phi_{ab}$
by $\sim$1 meV.  
We compare this to the system with $ab=$ (3.85 \AA)$^2$, 
$\sim$3\% tensile strain, 
and find that the uniaxial relaxation is no longer energetically favorable, and the
ground-state retains $\phi_{ab}$ rotations and $a=b$.

Our results indicate that LAO films on substrates with a lattice mismatch of
$\sim$0.5\% may, in theory, lower their energy through a partial relaxation such that 
$a \ne b$. The resulting structure exhibits $\phi_a$ rotations (for $a>b$) in
the space group $Fmmm$. In practice, no common growth substrate provides these
conditions, but the result is likely to be a general phenomenon in similar materials.
The data suggest an explanation for the experimental observation of $a\ne b$
in some coherent films grown on square substrates.
For example,  BiFeO$_3$ films thinner
than 50 nm grown on LAO substrates are reported to have $a\sim$3.84 \AA\ and $b\sim$3.76
\AA, compared to the substrate parameters $a=b=3.79$ \AA, indicating a uniaxial
relaxation of the epitaxial constraint.\cite{Zeches_et_al:2009}
Our results for LAO at 3\% tensile strain, which do not predict a uniaxial
strain relaxation, are consistent with experimental data
for films of LAO on STO (a mismatch strain of 3\%). Such films are reported to have square
in-plane lattice parameters within an accuracy of 0.01 \AA, indicating  that any
distortion of $a/b$ must be smaller than 0.5\%.\cite{Merckling_et_al:2007,
Reyren_et_al:2007,Huijben_et_al:2009}

\begin{table*}
\begin{tabular}{ m{3.0cm} m{1.2cm} m{1.2cm} m{1.2cm} m{1.5cm} m{1.7cm} m{1.5cm} m{1.5cm} m{1.2cm}}
\hline\hline\noalign{\smallskip}
 &  a ({\AA})  & b  ({\AA}) & c  ({\AA}) & $\beta$ &  Wyckoff pos. & {\em x } & {\em y } & {\em z } \\
\noalign{\smallskip}
\hline\noalign{\smallskip}

$C2/c$ & 9.15 & 5.2976 & 5.2976 & 54.927$^\circ$ &  La (4e) &  0.0 & 0.2500 & 0.25 \\
{\em (0\% strain)} &  &  &  &  &  Al (4d) & 0.25 &  0.25 & 0.5 \\
 &  &  &  &  &  O (8f) & -0.2652 &  0.4829 & 0.2823 \\
 &  &  &  &  &  O (4e) & 0.0 & -0.2804 & 0.25   \\
\noalign{\smallskip}
\hline\noalign{\smallskip}

$I4/mcm$&  5.2712 & 5.2712 & 7.5634 & - & La (4b)& 0.0 &  0.5 &  0.25   \\
 {\em(-0.5\% strain)} &  &  &  &  &  Al (4c)&   0.0&   0.0&   0.0\\
 &  &  &  &  &  O (8h) & -0.2212 &  0.2788 &  0.0   \\
 &  &  &  &  &  O (4a) &  0.0 &  0.0 &  0.25   \\

\noalign{\smallskip}
\hline\noalign{\smallskip}
$Imma$ &  7.4642 &  5.3156 &  5.3156 & - & La (4e) & 0.0 & 0.25 & -0.2491 \\ 
 {\em(0.5\% strain)} &  &  &  &  &  Al(4c) & 0.25& 0.25& 0.25\\ 
 &  &  &  &  &  O (8f) & -0.2298 & 0.0& 0.0\\ 
 &  &  &  &  &  O (4e) & 0.0& 0.25 & 0.2894 \\ 
\noalign{\smallskip}
\hline\noalign{\smallskip}

$Fmmm$ & 7.4800 & 7.4502 & 7.5099 & - & La (8i) & 0.0& 0.0&  -0.2501  \\
 {\em($\sim$0\% strain)} &  &  &  &  &  Al (8e) &   0.25& 0.25 & 0.0 \\
 &  &  &  &  &  O (8f) &   0.25& 0.25 & 0.25 \\
 &  &  &  &  &  O (8g) &  -0.2770  & 0.0 & 0.0 \\
 &  &  &  &  &  O (8h) &   0.0& 0.2231 & 0.0 \\
\noalign{\smallskip}
\hline\hline
\end{tabular}
\caption{Calculated structural parameters of each phase given for a representative value
of strain.  Note that here the labels $a, b, c$ do not refer to the pseudocubic lattice
parameters but to the conventional parameters of the given space group.
 }
\label{table2}
\end{table*}

\section{Disabled octahedral rotations}

Finally, we address the effect of manually disabling octahedral rotations.  
For this we treat a five-atom tetragonal unit cell in which
rotation of the oxygen octahedra is forbidden by symmetry. 
 At each value of epitaxial strain we
relax the positions of the ions and calculate the electric polarization using the Berry
phase method.\cite{King-Smith/Vanderbilt:1993,Vanderbilt/King-Smith:1993}  
We find that, while the unstrained system is non-polar, an abrupt
transition to a state with large polarization occurs at -3\% strain. The polarization is
out-of-plane and results from a structural distortion within $P4/mmm$ symmetry
 in which Al and La displace along [001]. 
A compressive strain of 4\%
results in a polarization of 38 $\mu$C cm$^{-2}$ relative to a centrosymmetric
reference structure.  
These results indicate the existence of an incipient ferroelectric mode that is usually
suppressed by the dominant antiferrodistortive rotational modes.
%

\section{Summary and discussion}

In summary, we find that several different phases of LAO, characterized by distinct
patterns of octahedral rotations, can be stabilized by varying the epitaxial
constraints over a range that is readily accessible experimentally.   
This remarkable structural softness indicates that it is unlikely that the bulk-type
 $\phi_{abc}$ rotations will be observed in heteroepitaxial thin films of LAO on
any substrate.  This has important implications for the interpretation of structural
data for LAO films, with particular relevance to investigations of the properties of
LAO/STO interfaces.

We find that biaxial strain of -0.2\% to 0.1\% results in large changes
in octahedral tilt angles in the bulk-like $C2/c$ phase of LAO.  
Larger strains, however, induce a tilt-driven transition to one of three phases in which
at least one tilt component is zero ($I4/mcm$, $Imma$, or $Fmmm$).
Within these strained phases, we find that changes in lattice parameters
related to epitaxial strain are largely accommodated by changes in the Al-O bond
lengths, with only small changes to tilt angles, and without any significant effect on
electronic properties.  
These findings provide insight to the growing list of complex oxide perovskites in which
the rotational distortions are observed to depend on lattice parameters.
Given the coupling between rotations and strain in LAO, and its non-polar
structure,  we propose it as a model system for exploring new non-linear optical
techniques for probing octahedral rotations.\cite{Denev_et_al:2008}

Finally, for small values of tensile strain we find that a uniaxial relaxation of the 
strain, such that $a> b$, stabilizes $\phi_a$ rotations over $\phi_{ab}$.  
Our results suggest a route to selectively stabilize different tilt patterns via the  
substrate geometry. Conversely, these results also suggest an explanation for the
unequal in-plane lattice parameters observed in some epitaxial thin films of
perovskites, in which similar energetics of  $\phi_a$  and $\phi_{ab}$ rotations may
drive a distortion of $a/b$.

\begin{acknowledgments}

This work was supported by the National Science Foundation (NSF), Grants No. NIRT-0609377 (NAS) and DMR-0820404 (AJH).  Computational support was
provided by the MRL Central Facilities under the NSF MRSEC Program, Award No.
DMR05-20415, and the CNSI Computer Facilities at UC Santa Barbara, NSF Grant No.
CHE-0321368.  We thank J.~M.~Rondinelli and M.~Stengel for helpful discussions.
\end{acknowledgments}

\bibliographystyle{apsrev}


\end{document}